\documentstyle[aps,prb,psfig,preprint,tighten,floats]{revtex}
\begin{document} 
\draft 

\title{Hierarchical model for the scale-dependent velocity\\
of seismic waves} 
\author{\sf J. TWORZYD\L O$^{\rm a,b}$ \& C. W. J. BEENAKKER$^{\rm a}$} 
\address{\small\sf $^{\rm a}$Instituut-Lorentz, Universiteit Leiden,
P.O. Box 9506, 2300 RA Leiden, The Netherlands\\
$^{\rm b}$Institute of Theoretical Physics, Warsaw University,
Hoza 69, 00-681 Warszawa, Poland}
\maketitle

\begin{figure} 
\centerline{\psfig{figure=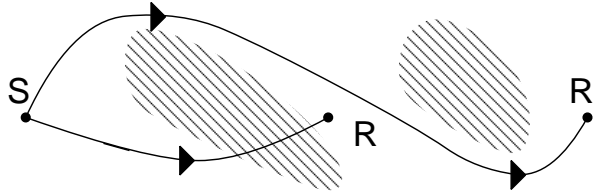,width=8cm}\medskip}
\caption[]{ Illustration of the scale dependent velocity.
Two rays are shown of short wavelength waves emitted from a source
$S$ and recorded at two receivers $R$, $R'$. The shaded areas
indicate regions of slow propagation. Each ray follows the path
of least time from source to receiver, in accordance with
Fermat's principle. The longer 
trajectory seeks out the fastest path more efficiently
than the shorter one, hence the apparent increase in velocity.
Perturbation theory breaks down when the deviation of the ray from
the straight path becomes comparable with the characteristic
size of the heterogeneities.
\label{fermat_plot}}
\end{figure}

\noindent
{\bf
Elastic waves of short wavelength propagating through the 
upper layer of the Earth appear to move faster at large
separations of source and receiver than at short separations.
This scale dependent velocity is a manifestation of Fermat's
principle of least time in a medium with random velocity
fluctuations. Existing perturbation theories predict a linear
increase of the velocity shift with increasing separation,
and cannot describe the saturation of the velocity shift at
large separations that is seen in computer simulations. Here
we show that this long-standing problem in seismology can be 
solved using a model developed originally in the context of
polymer physics. We find that the saturation velocity scales
with the four-third power of the root-mean-square amplitude
of the velocity fluctuations, in good agreement with the 
computer simulations.
}\bigskip\\
Seismologists probe the internal structure of the Earth by recording
the arrival times of waves created by a distant earthquake or
explosion\cite{Cla85}.  Systematic differences between studies based on
long and short wavelengths $\lambda$ have been explained\cite{Nol93}
in terms of a {\em scale dependence\/} of the velocity at short
wavelengths. The velocity obtained by dividing the separation $L$ of
source and receiver by the travel time $T$ increases with increasing
$L$, because --- following Fermat's principle --- the wave seeks out
the fastest path through the medium (see Fig.\ \ref{fermat_plot}). This
search for an optimal path is more effective for large separations,
hence the apparent increase in velocity on long length scales. It
is a short-wavelength effect, since Fermat's principle breaks
down if the width $\sqrt{L\lambda}$ of the first Fresnel zone
becomes comparable to the size $a$ of the heterogeneities. The
scale-dependent velocity of seismic waves was noted by Wielandt
more than a decade ago\cite{Wie87}, and has been studied extensively by
geophysicists\cite{Pet90,Mul92,Rot93,Ave94,Muk95,Wit96,Sha96,Rot97,Bri97,Ton98,Sam98,Sam98b}.

A rather complete solution of the problem for small $L$ was given by
Roth, M\"{u}ller, and Snieder\cite{Rot93}, by means of a perturbation
expansion around the straight path. The velocity shift $\delta
v=v_0(1- v_{0} T/L)$ (with $v_{0}$ the velocity along the straight
path) was averaged over spatially fluctuating velocity perturbations
with a Gaussian correlation function (having correlation length $a$
and variance $\varepsilon^{2}v_{0}^{2}$, with $\varepsilon\ll 1$). It
was found that $\langle \delta v\rangle\simeq v_{0}\varepsilon^{2}L/a$
increases linearly with $L$.  Clearly, this increase in velocity can not
continue indefinitely. The perturbation theory should break down when
the root-mean-square deviation $\delta x\simeq\varepsilon a(L/a)^{3/2}$
of the fastest path from the straight path becomes comparable to
$a$. Numerical simulations\cite{Rot93,Wit96,Sam98} show that the velocity
shift saturates on length scales greater than the critical length $L_{\rm
c}\simeq a\varepsilon^{-2/3}$ for the validity of perturbation theory. A
theory for this saturation does not exist.

It is the purpose of this article to present a {\em non-perturbative\/}
theory for this seismological problem, by making the analogy with a
problem from polymer physics.  The problem of the velocity shift in a
random medium belongs to the class of optimal path problems that has
a formal equivalence to the directed polymer problem\cite{Hal95}. The
mapping between these two problems relates a wave propagating through a
medium with velocity fluctuations to a polymer moving in a medium with
fluctuations in pinning energy. The travel time of the wave between source
and receiver corresponds to the energy of the polymer with fixed end
points. At zero temperature the configuration of the polymer corresponds
to the path selected by Fermat's principle. (The restriction to directed
polymers, those which do not turn backwards, becomes important for higher
temperatures.) There exists a simple solvable model for directed polymers,
due to Derrida and Griffiths\cite{Der89}, that has remained unnoticed
in the seismological context. Using that model we can go beyond the
breakdown of perturbation theory and describe the saturation of the
velocity shift on large length scales.

\begin{figure} 
\centerline{\psfig{figure=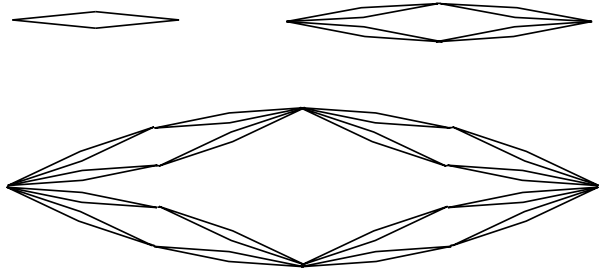,width=8cm}\medskip}
\caption[]
{First three steps of the recursive 
construction of the hierarchical lattice.
\label{lattice}} 
\end{figure}

{\bf Hierarchical model}\\
We follow a recursive procedure, by which the probability distribution
of travel times is constructed at larger and larger distances,
starting from the perturbative result at short distances. At each
iteration we compare travel times from source to receiver along two
branches, choosing the smallest time.  A branch consists of two bonds,
each bond representing the length scale of the previous step. This
recursive procedure produces the lattice of Fig.\ \ref{lattice},
called a hierarchical lattice\cite{Der89}. The lattice in this example
represents a two-dimensional system, since at each step the length is
doubled while the number of bonds is increased by a factor of four. For
the three-dimensional version one would compare four branches at each
step (each branch containing two bonds), so that the total number of
bonds would grow as the third power of the length. Since most of the
simulations have been done for two-dimensional systems, we will consider
that case in what follows.

To cast this procedure in the form of a recursion relation, we
denote by $p_{k}(T)$ the distribution of travel times at step
$k$. One branch, consisting of two bonds in series, has travel
time distribution
\begin{equation} 
q_{k}(T)=\int_{0}^{\infty}dT'\,p_{k}(T')p_{k}(T-T'),\label{qkdef}
\end{equation} 
assuming that different bonds have uncorrelated distributions. To
get the probability distribution at step $k+1$ we compare travel
times of two branches,
\begin{eqnarray} 
p_{k+1}(T)&=& \int_{0}^{\infty}dT'\int_{0}^{\infty}dT''\,
\delta\biglb(T-{\rm min}(T',T'')\bigrb) q_k(T')
q_k(T'')\nonumber\\
&=&2q_{k}(T)\int_{T}^{\infty}dT'\,q_{k}(T').\label{pkdef}
\end{eqnarray}
We start the recursion relation at step $0$ with the distribution
$p_{0}(T)$ calculated from perturbation theory at length $L_{c}$.
Iteration of eq.\ (\ref{pkdef}) then produces the travel
time distribution $p_{k}(T)$ at length $L=2^{k}L_{c}$.

Equation (\ref{pkdef}) is a rather complicated non-linear
integral equation. Fortunately, it has several simplifying
properties\cite{Der89,Rou91}. One can separate out the mean $\left<
T \right>_0$ and standard deviation $\sigma_0\neq 0$ of the
starting probability distribution, by means of the $k$-dependent
rescaling $\tau = (T-2^k \left< T\right>_0)/\sigma_0$. The recursion
relation (\ref{pkdef}) is invariant under this rescaling, which
means that we can restrict ourselves to starting distributions
$\tilde{p}_0(\tau)=\sigma_0p_0(\sigma_0 \tau+\left< T \right>_0)$ having
zero mean and unit variance. This is the first simplification. After
$k$ iterations the mean $\tilde{m}_k$ and standard deviation
$\tilde{\sigma}_k$ of the rescaled distribution $\tilde{p}_k(\tau)$
yield the mean $\left<T\right>_k$ and standard deviation $\sigma_k$
of the original $p_k(T)$ by means of
\begin{equation}
\left<T\right>_k = \sigma_0 \tilde{m}_k + 2^k\left<T\right>_0,\
\sigma_k = \sigma_0 \tilde{\sigma}_k. \label{mk2tk}
\end{equation}
The second simplification is that for large $k$, the recursion relation
for $\tilde{p}_k(\tau)$ reduces to\cite{Der89,Rou91}
\begin{equation}
\tilde{p}_{k+1}(\tau)= \textstyle{\frac{1}{2}}\alpha
             \tilde{p}_k\left( \textstyle{\frac{1}{2}} \alpha \tau
                        + \beta \tilde{\sigma}_k  
                        + (1-\alpha)\tilde{m}_k \right),
\label{map}                        
\end{equation}
with universal constants $\alpha=1.627$ and $\beta=0.647$.
Under the mapping (\ref{map}), the mean and standard deviation 
evolve according to
\begin{equation}
\tilde{m}_{k+1} = 2 \tilde{m}_{k} - 2 \beta \tilde{\sigma}_k/\alpha ,\ 
\tilde{\sigma}_{k+1} = 2 \tilde{\sigma}_k/\alpha.
\label{rec}
\end{equation}
The solution of this simplified recursion relation is
\begin{equation}
\tilde{m}_{k} = \frac{2^k \beta}{\alpha-1} 
                ( A  \alpha^{-k} - B),\ 
\tilde{\sigma}_{k} = 2^k A \alpha^{-k}.
 \label{mksol}
\end{equation}
The coefficients $A$ and $B$ are non-universal, depending on the shape
of the starting distribution $\tilde{p}_0$. For a Gaussian $\tilde{p}_0$
we find $A=0.90$, $B=0.95$, close to the values $A=1$, $B=1$ that would
apply if eq.\ (\ref{rec}) holds down to $k=0$. For a highly distorted
bimodal $\tilde{p}_0$ we find $A=0.71$, $B=0.88$.  We conclude that $A$
and $B$ depend only weakly on the shape of the starting distribution.

{\bf Scaling laws}\\
Given the result (\ref{mksol}) we return to the mean and standard
deviation of $p_k(T)$ using eq.\ (\ref{mk2tk}).  Substituting
$k={\rm log}_{2}(L/L_c)$ one finds the large-$L$ scaling laws
\begin{eqnarray}
\frac{\left< T \right>}{L} &= & \frac{\left< T \right>_0}{L_c} -
\frac{\beta}{\alpha-1} \frac{\sigma_0}{L_c}
\left[B-A\left(\frac{L_c}{L}\right)^{p}\right],
\label{sol1}\\
\frac{\sigma}{L} &=& \frac{\sigma_0}{L_c} A
\left(\frac{L_c}{L}\right)^{p}.
\label{sol2}
\end{eqnarray}
The mean travel time $\left<T\right>$ and standard deviation $\sigma$
scale with $L$ with an exponent $p={\rm log}_{2}\alpha=0.702$.  This
scaling exponent has been studied intensively for the directed polymer
problem\cite{Hal95}.

For the seismic problem the primary interest is not the
scaling with $L$, but the scaling with the strength $\varepsilon$
of the fluctuations. Perturbation theory\cite{Rot93} gives the
$\varepsilon$-dependence at length $L_c$,
\begin{equation}
1-v_0 \left< T \right>_0/L_c \simeq \varepsilon^2 L_c/a,\ 
v_0 \sigma_0/L_c \simeq \varepsilon \sqrt{a/L_c},
\label{per1}
\end{equation}
where $\simeq$ indicates that coefficients of order unity
have been omitted. (We will fill these in later.)
Since $L_c\simeq a \varepsilon^{-2/3}$ (as mentioned in 
the introduction), we find upon substitution into
eq.\ (\ref{sol1}) the scaling of the mean velocity shift
at length $L\gg L_c$:
\begin{equation}
\left< \delta v \right>/v_0 \equiv
1-v_0 \left<T\right>/ L \simeq \varepsilon^{4/3} \left[1+{\cal O}(L_c/L)^p\right].
\end{equation}
The mean velocity shift saturates at a value of order 
$v_0 \varepsilon^{4/3}$.
The exponent $\frac{4}{3}$ was anticipated in ref.\ \onlinecite{Sam98}
and is close to the value $1.33\pm 0.01$ resulting from
simulations\cite{Rot93}.

{\bf Comparison with simulations}\\
For a more quantitative description we need to know the
coefficients omitted in eq.\ (\ref{per1}). These
are model dependent\cite{Rot93,Muk95,Sha96}. 
To make contact with the
simulations\cite{Rot93,Wit96} we consider
the case of an incoming plane wave instead of a point source.
The perturbation theory for the mean velocity shift at length
$L_c$ gives
\begin{equation}
\left< \delta v \right>_0= v_0 \varepsilon^2\frac{L_c}{a}
 \frac{\sqrt{\pi}}{2}
 \left(1-\frac{2}{\sqrt{\pi}}\frac{a}{L_c}\right).
 \label{vper}
\end{equation}
The variance at length $L_c$ is
\begin{equation}
\left< \delta v^2 \right>_0= v_0^2 \varepsilon^2\frac{a}{L_c}
 \sqrt{\pi}
 \left(1 -\sqrt{\pi}\varepsilon^2\frac{L_c^3}{a^3}\right).
 \label{dvper}
\end{equation}
We quantify the criterion for the breakdown of perturbation
theory by 
$L_c=\kappa a \varepsilon^{-2/3}$, with $\kappa=0.765$ our
single fit parameter. For the non-universal constants
$A$ and $B$ we can use in good  approximation $A=1$, $B=1$.
The mean velocity shift
in the non-perturbative regime ($L>L_c$) is then expressed as:
\begin{equation}
\left<\delta v\right> =  \frac{\beta}{\alpha-1} 
                         \sqrt{\left< \delta v^2\right>_0}
                   \left[ 1-\left( \frac{L_c}{L} \right)^p \right]
                         + \left< \delta v \right>_0.
\label{vnonper}
\end{equation}
For $L<L_c$ we use the perturbative result (\ref{vper})
(with $L_c$ replaced by $L$). 
As shown in Fig.\ 3, the agreement with the computer simulations
is quite satisfactory, in particular in view of the fact that there is
a single fit parameter $\kappa$ for all curves.

\begin{figure} 
\centerline{\psfig{figure=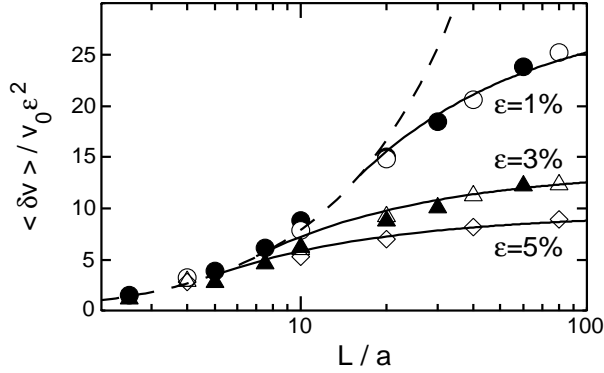,width=8cm}\medskip}
\caption[]
{Scale dependence of the velocity, showing the saturation of the mean
velocity shift $\langle\delta v\rangle$ at large separations $L$ of source
and receiver. The dashed curve is the result (\ref{vper}) of perturbation
theory, the solid curves are the non-perturbative results from eq.\
(\protect\ref{vnonper}).  Data points are results of computer simulations
at various strengths $\varepsilon$ of the velocity perturbation (open
markers from ref.\ \protect\onlinecite{Rot93}, filled markers from ref.\
\protect\onlinecite{Wit96}).
\label{datapoints}}
\end{figure}

{\bf Conclusion}\\
We have presented a non-perturbative theory of the scale-dependent
seismic velocity in heterogeneous media. The saturation of the velocity
shift at large length scales, observed in computer simulations, is well
described by the hierarchical model --- including the $\varepsilon^{4/3}$
scaling of the saturation velocity. We have concentrated on the case
of two-dimensional propagation (for comparison with the simulations),
but the $\varepsilon^{4/3}$ scaling holds in three dimensions as well. (The
coefficients $\alpha=1.74$, $\beta=1.30$ are different in 3D.)

Our solution of the seismic problem relies on the mapping onto the
problem of directed polymers. This mapping holds in the short-wavelength
limit, $\lambda\lesssim a^{2}/L$. To observe the saturation at $L\simeq
L_{c}$ thus requires $\lambda\lesssim a\varepsilon^{2/3}$. For $L\gtrsim
a^{2}/\lambda$ the velocity shift will decrease because the velocity
fluctuations are averaged out over a Fresnel zone. There exists a
perturbation theory\cite{Sam98b} for the velocity shift that includes
the effects of a finite wavelength. It is a challenging problem to see
if these effects can be included into the non-perturbative hierarchical
model as well.

{\bf Acknowledgements.}
We are indebted to R. Snieder for suggesting this problem to us,
and to X. Leyronas for valuable discussions throughout this
work. We acknowledge the support of the Dutch Science Foundation
NWO/FOM.



\begin{references}

\bibitem{Cla85} Claerbout, J. F. {\em Imaging the Earth's
Interior\/} (Blackwell, Oxford, 1985).
\bibitem{Nol93} Nolet, G. \& Moser, T. J. Teleseismic delay times in a 3-D Earth
and a new look at the $S$ discrepancy {\em Geophys. J. Int.\/} {\bf
114}, 185--195 (1993).
\bibitem{Wie87} Wielandt, E. On the validity of the ray approximation for
interpreting delay times. In {\em Seismic Tomography} (ed.
Nolet, G.) 85--98 (Reidel, Dordrecht, 1987).
\bibitem{Pet90} Petersen, N. V. Inverse kinematic problem for a random gradient
medium in geometric optics approximation. {\em Pure Appl.\ Geophys.\/} {\bf 132},
417--437 (1990).
\bibitem{Mul92} M\"{u}ller, G., Roth, M. \& Korn, M. Seismic-wave traveltimes in
random media. {\em Geophys.\
J. Int.\/} {\bf 110}, 29--41 (1992).
\bibitem{Rot93} Roth, M., M\"{u}ller, G. \& Snieder, R. Velocity shift in random
media. {\em Geophys.\
J. Int.\/} {\bf 115}, 552--563 (1993).
\bibitem{Ave94} Van Avendonk, H. \& Snieder, R. A new mechanism for shape induced seismic
anisotropy. {\em Wave Motion} {\bf
20}, 89--98 (1994).
\bibitem{Muk95} Mukerji, T., Mavko, G., Mujica, D. \& Lucet, N.
Scale-dependent seismic
velocity in heterogeneous media.
{\em Geophys.\/} {\bf 60}, 1222--1233 (1995).
\bibitem{Wit96} Witte, O., Roth, M. \& M\"{u}ller, G. Ray tracing in random
media. {\em Geophys.\
J. Int.\/} {\bf 124}, 159--169 (1996).
\bibitem{Sha96} Shapiro, S. A., Schwarz, R. \& Gold, N. The effect of random
isotropic inhomogeneities on the phase velocity of seismic
waves. {\em Geophys.\
J. Int.\/} {\bf 127}, 783--794 (1996).
\bibitem{Rot97} Roth, M. Statistical interpretation of traveltime
fluctuations. {\em Phys.\ Earth Planet.\ Inter.\/} {\bf 104},
213--228 (1997).
\bibitem{Bri97} Brittan, J. \& Warner, M. Wide-angle seismic velocities in
heterogeneous crust. {\em Geophys.\ J. Int.\/} {\bf
129}, 269--280 (1997).
\bibitem{Ton98} Tong, J., Dahlen, F. A., Nolet, G. \&
Marquering, H. Diffraction effects upon finite-frequency travel times: a
simple 2-D example. {\em Geophys.\ Res.\ Lett.\/} {\bf 25}, 1983--1986 (1998).
\bibitem{Sam98} Samuelides, Y. \& Mukerji, T. Velocity shift in heterogeneous media
with anisotropic spatial correlation. {\em Geophys.\ J. Int.\/}
{\bf 134}, 778--786 (1998).
\bibitem{Sam98b} Samuelides, Y. Velocity shift using the Rytov
approximation. {\em J. Acoust. Soc. Am.\/} {\bf 104},
2596--2603 (1998).
\bibitem{Hal95} Halpin-Healy, T. \& Zhang, Y. C. Kinetic roughening phenomena, stochastic
growth, directed polymers and all that. {\em Phys.\ Rep.\/}
{\bf 254}, 215--415 (1995).
\bibitem{Der89} Derrida, B. \& Griffiths, R. B. Directed polymers on disordered
hierarchical lattices. {\em Europhys.\ Lett.\/}
{\bf 8}, 111--116 (1989).
\bibitem{Rou91} Roux, S., Hansen, A., Da Silva, L. R., Lucena, L. S. \&
Pandey, R. B. Minimal path on the hierarchical diamond lattice. {\em J. Stat.\ Phys.\/}
{\bf 65}, 183--204 (1991).

\end{references}
\end{document}